\documentclass[showpacs,twocolumn]{svjour3}
\usepackage[latin1]{inputenc}
\usepackage[english]{babel}
\usepackage[T1]{fontenc}

\usepackage{graphicx}
\usepackage{amsmath, amssymb}
\usepackage{dcolumn}
\usepackage{bm}
\usepackage{subfigure}

\newcommand{\be}{\begin{equation}}
\newcommand{\ee}{\end{equation}}
\newcommand{\bq}{\begin{eqnarray}}
\newcommand{\eq}{\end{eqnarray}}

\journalname{International Journal of Theoretical Physics}

\def\ca{\c{c}\~{a}}

\begin{document}

\title{\sc Naturalness and theoretical constraints on the Higgs boson mass
}

\author{ A. R. Vieira         \and
        Brigitte Hiller \and
M. C. Nemes \and Marcos Sampaio 
}

\institute {A. R. Vieira \at Federal University of Minas Gerais, Physics Department, ICEx, PO Box 702, 30.161-970, Belo Horizonte MG, Brazil\\
\email{arvieira@fisica.ufmg.br} \and Brigitte Hiller \at Coimbra University - Faculty of Science and Technology- Physics Department - Center of Computational Physics,
Rua Larga, P-3004-516 Coimbra - Portugal\\ \email{brigitte@teor.fis.uc.pt} \and M. C. Nemes 
\at Federal University of Minas Gerais, Physics Department, ICEx, PO Box 702, 30.161-970, Belo Horizonte MG, Brazil\\ \email{carolina@fisica.ufmg.br}
\and Marcos Sampaio 
\at Federal University of Minas Gerais, Physics Department, ICEx, PO Box 702, 30.161-970, Belo Horizonte MG, Brazil\\ \email{msampaio@fisica.ufmg.br}
}
\date{Received: date / Accepted: date}

\maketitle

\begin{abstract}

Arbitrary regularization dependent parameters in Quantum Field Theory are usually fixed on symmetry or phenomenology grounds. We verify that the quadratically divergent behavior responsible for the lack of naturalness in the Standard Model (SM) is intrinsically arbitrary and regularization dependent. While quadratic divergences are welcome for instance in effective models of low energy QCD, they pose a problem in the SM treated as an effective theory in the Higgs sector. Being the very existence of quadratic divergences a matter of debate, a plausible scenario is to search for a symmetry requirement that could fix the arbitrary coefficient of the leading quadratic behavior to the Higgs boson mass to zero. We show that this is possible employing consistency of scale symmetry breaking by quantum corrections. Besides eliminating a fine-tuning problem and restoring validity of perturbation theory, this requirement allows to construct bounds for the Higgs boson mass in terms of $\delta m^2/m^2_H$ (where $m_H$ is the renormalized Higgs mass and $\delta m^2$ is the 1-loop Higgs mass correction). Whereas $\delta m^2/m^2_H<1$ (perturbative regime) in this scenario allows the Higgs boson mass around the current accepted value, the inclusion of the quadratic divergence demands $\delta m^2/m^2_H$ arbitrarily large to reach that experimental value.

\end{abstract}

\section{Introduction}\label{s1}
The Higgs field in the Standard Model (SM) is fundamental to ensure its renormalizability and unitarity. 
However, as pointed out by Susskind \cite{Susskind}, there are self-energy corrections to the
Higgs propagator which would require an incredible fine-tuning of $10^{-34}$ parts in $1$,
if we assume that the limit of validity of the SM is at the Planck scale. This problem does not
appear in other theories which present {\it Naturalness}. The concept of Naturalness was defined by Susskind as when 
{\it the behavior of the world at ordinary energies is not exceedingly sensitive to the values of the fundamental parameters} \cite{Susskind2}.
Theories with scalar fields present unnaturalness because the mass of the scalar, a phenomenological parameter, exceedingly depends on the 
cutoff of the theory. In the context of the Standard Model, we can see this,
after a regularization and a renormalization scheme, via the corrected mass of the Higgs boson \cite{{Veltman},{Castro}}
\begin{align}
  m ^ 2_H =& m^2+\frac{3 \Lambda^2}{8\pi ^2 \upsilon^2}[m_{Z}^2+2m_{W}^2+m^2-4m_{t}^2]+ \nonumber\\
 &+ O(\ln\frac{\Lambda}{m})
  \label{1}
\end{align}
where $m_{Z}$, $m_{W}$, $m$ and $m_{t}$ stand for $Z^{0}$, $W^{\pm}$, Higgs and top quark masses, respectively, and $\Lambda$ is the energy scale of 
the theory. We see that large corrections, unnaturalness and the consequent fine-tuning problem are all caused by $\delta m^2=m^2_H-m^2$ being 
proportional to the energy scale squared. 

Therefore, proposals to eliminate this quadratic energy scale have been put forth \cite {Veltman,Bardeen,Grange,Aoki}. One of them consisted in choosing the sum in square brackets in eq. (\ref {1}) to be zero \cite {Veltman}. This gives a constraint among the masses. However, its validity at any energy scale is not guaranteed \cite{{Bardeen},{Chaichian}}. Such constraint overestimates the Higgs boson mass \cite {Castro} as compared with the expected value \cite{Higgs}. 

In ref. \cite{Bardeen} is argued that the $\Lambda^2$ dependence of equation (\ref{1}) is not in consonance with scale symmetry breaking. In order to see this consider the classical energy-momentum tensor which is broken by the mass term
\begin{equation}
 \Theta_{\mu}^{\mu}\mid_{classical} = m^2 \bar{H} H.
\label{e1}
\end{equation}
where $H$ denotes the Higgs field.

We have a symmetry in equation (\ref{e1}) in the limit $m\rightarrow 0$. The quantum corrections also break that conservation law:
\begin {eqnarray}
\centering
  &\Theta_{\mu}^{\mu}\mid_{1-loop} = \delta m^2 \bar{H} H + \sum_i \frac{\partial \mathcal{L}}{\partial \lambda_i} \beta_{\lambda_i}
 \nonumber\\
 &\delta m^2= \frac{3 \Lambda^2}{8\pi ^2 \upsilon^2}[m_{Z}^2+2m_{W}^2+m^2-4m_{t}^2] + \nonumber\\
&+ O(\ln\frac{\Lambda}{m})
  \label{e2}
\end{eqnarray}
where $\beta_{\lambda_i}$ is the beta function associated with the coupling $\lambda_i$. 

If we try to restore the classical limit, taking the limits $m\rightarrow 0$ and $\beta_{\lambda_i}\rightarrow 0$ in equation (\ref{e2}), the result is
\begin {eqnarray}
\centering
  &\Theta_{\mu}^{\mu}\mid_{1-loop} = [\frac{3 \Lambda^2}{8\pi ^2 \upsilon^2}(m_{Z}^2+2m_{W}^2-4m_{t}^2)]\bar{H} H
  \label{e3}
\end{eqnarray}

Equation (\ref{e3}) shows that we do not restore the classical limit ($\Theta_{\mu}^{\mu}=0$) when we take the classical 
limit ($m\rightarrow 0$ and $\beta_{\lambda_i}\rightarrow 0$). In the limit where $m$ goes to zero, in order to preserve the 
structure of the anomalous divergence of the scale current, $\delta m^2$ must scale as $m^2$ rather than $\Lambda^2$. The 
conclusion is that this $\Lambda^2$ dependence  is incompatible with the consistency of scale symmetry breaking,  which leads 
us to analyze the possibility and implications of it being an artifact of regularization.  In fact in eq. (\ref{e2}) the 
$\Lambda^2$ dependence results from the use of a sharp cutoff regularization, which breaks scale symmetry
but not in a physical way, i.e. not related with the beta functions. The scheme causes spurious divergences.

A similar conclusion is drawn in ref. \cite{Grange}, using the Taylor-Lagrange renormalization. Furthermore in the Wilsonian renormalization group approach it was argued in ref. \cite{Aoki} that the consistency of scale invariance breaking of the SM is an alternative solution to the Naturalness problem.

In this work we show in a regularization independent way that the coefficient of the quadratic divergence which gives rise to the Naturalness problem is intrinsically arbitrary and should be set to zero on the grounds of the symmetry arguments discussed above. This is in consonance with the approach proposed in ref. \cite{Jackiw}, in which arbitrary parameters stemming from perturbation theory calculations should be fixed by the underlying symmetries of the model. 

As we will explicitly show in the next section, a quadratic divergence can be generally parametrized as $I_{quad}(\mu ^2)=\frac {i}{(4\pi )^2}[c_{2}\Lambda^ 2 +(1+c_{1})\mu ^2+\mu^ 2 \ln \frac{\Lambda ^2}{\mu ^2}]$, where $c_2$ is an arbitrary regularization
dependent constant.

Moreover, after these considerations, we can reestablish perturbation theory choosing the ratio $\delta m^2/m^2_H$ in a range between $0$ to $1$.
We obtain bounds for the Higgs boson mass in a $m_H$ {\it versus} $\Lambda$ diagram. We compare our bounds with those existent in the literature \cite {{Regiao2},{Regiao1},{Regiao3}}.

\section{Implicit Regularization and arbitrariness in divergent amplitudes}\label {s2}

An ultraviolet (UV) divergent amplitude will be regularized implicitly \cite{IR}. In other words, we assume the existence of a regulator function
for the integral and we separate the regularization dependent content from the finite one by using the following identity
\begin{align}
 \frac{1}{[(k-l)^2-\mu ^2]}&=\sum_{j=0}^{n-1}\frac{(-1)^j(l^2-2 l.k)^j}{(k^2-\mu ^2)^{j+1}}+ \nonumber\\
 &+ \frac{(-1)^n(l^2-2 l.k)^n}{(k^2-\mu ^2)^n[(k-l)^2-\mu ^2]}
 \label{2}
\end{align} 
where $k$ is the internal momentum, $\mu$ is the mass of the internal particle and $n$ is chosen such that the denominator of an UV divergent integral does not contain an external momentum $l$. We get basic divergent integrals which are defined as
\begin{equation}
 I_{quad}(\mu ^2)\equiv \int^{\Lambda} \frac{d^4 k}{(2\pi)^4} \frac{1}{(k^2-\mu ^2)}
 \label{3}
\end{equation}
\begin{equation}
 I_{log}(\mu ^2)\equiv \int^{\Lambda} \frac{d^4 k}{(2\pi)^4} \frac{1}{(k^2-\mu ^2)^2}
 \label{4}
\end{equation}
where the index $\Lambda$ means that we assume the existence of a regulator just in order to perform mathematical manipulations with the 
integrand.

\begin{figure}[!htb]
 \centering
 \includegraphics[scale=0.55]{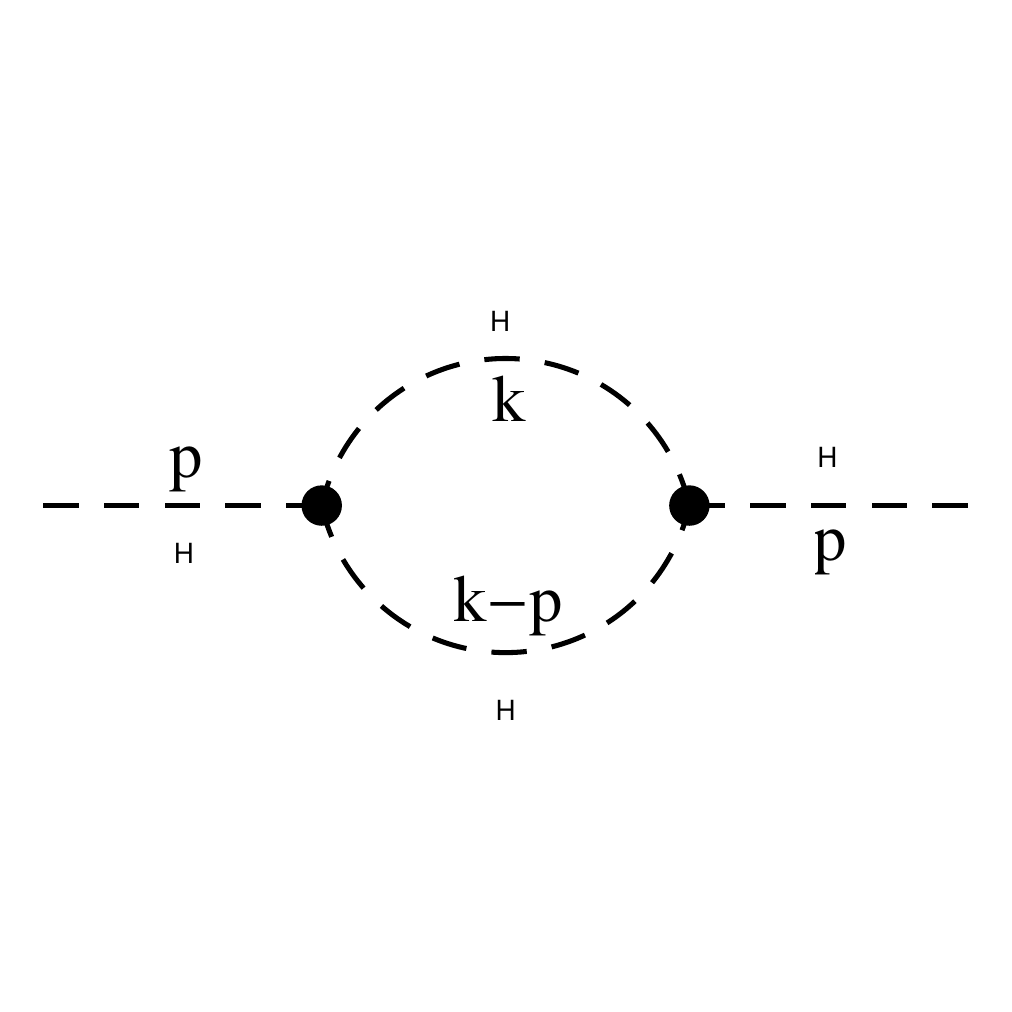}
 \caption{One of the corrections to the Higgs propagator.}
 \label{fig1}
\end{figure}

For example, one of the corrections to the Higgs propagator (see figure \ref {fig1}) is rewritten using eq. (\ref {2}) for $n=1$
 
\begin{align}
 &\int^{\Lambda}\frac{d^4 k}{(2\pi)^4} \frac{1}{(k^2-m ^2)[(k-p)^2-m^2]}= I_{log}(m^2) +\nonumber \\
 &+\int^{\Lambda}\frac{d^4 k}{(2\pi)^4} \frac{p^2-2 p.k}{(k^2-m^2)^2[(k-p)^2-m ^2]}
  \label{5}
\end{align}
The second term in (\ref {5}) is finite in the UV limit by power counting of the internal momentum.

The most general parametrization of the basic divergent integral $ I_{log}(\mu ^2)$ can be constructed as follows. By noting that
\begin{equation}
 \frac{\partial I_{log}(\mu ^2)}{\partial \mu ^2}=\frac {-i}{(4\pi )^2 \mu^2}
 \label{6}
\end{equation}
and
\begin{equation}
 \frac{\partial I_{quad}(\mu ^2)}{\partial \mu ^2}=I_{log}(\mu ^2),
 \label{7}
\end{equation}
should be obeyed by any regularization scheme, a general parametrization with explicit scale dependence that satisfies (\ref{6}) and (\ref{7}) is given by
\begin{equation}
 I_{log}(\mu ^2)=\frac {i}{(4\pi )^2}[\ln \frac{\Lambda ^2}{\mu ^2}+c_{1}]
 \label{8}
\end{equation}
\begin{equation}
 I_{quad}(\mu ^2)=\frac {i}{(4\pi )^2}[c_{2}\Lambda^ 2 +(1+c_{1})\mu ^2+\mu^ 2 \ln \frac{\Lambda ^2}{\mu ^2}],
 \label{9}
\end{equation}
where $c_1$ and $c_2$ are dimensionless and arbitrary constants.
Such arbitrary and regularization dependent constants are inherent to perturbation theory and can be fixed on symmetry grounds \cite{Jackiw}. For example, gauge 
and momentum routing invariance in QED demands that quadratic surface terms, which are related to arbitrary constants like $c_1$ and $c_2$, should vanish \cite {Adriano}.There are also some examples of extended QED where gauge invariance is used to fix Lorentz-violating terms \cite {Gustavo}.

A plethora of regularization schemes have been constructed to be used where gauge invariant Dimensional Regularization may fail, namely in the so called dimensional specific theories among which super-symmetric, chiral and topological quantum field theories figure in. A natural question would be which basic properties should a method that does not resort to analytical continuation in the space-time dimension should retain in order to be invariant. We start by illustrating with  simple examples following \cite{Varin:2006de}. Let $\Delta$ be the superficial degree of divergence of a $1$-loop integral where the momentum $k$ runs. Consider the following $\Delta=2$ integrals,
\be
A = \int_k \frac{k^2}{(k^2-\mu ^2)^2},
\ee
and
\be
B = I_{quad} (\mu ^2) + \mu ^2 I_{log} (\mu ^2),
\ee
where $\int_k \equiv \int d^4k/(2 \pi)^4$ and we recover the standard notation of eqs. \ref{3} and \ref{4}.

We expect  $A=B$ be guaranteed by any regularization procedure. However this is not the case. Proper-time regularization \cite{ZinnJustin:1993wc}, for instance, introduces a cut-off $\Lambda$ after Wick rotation via the following identity at the level of propagators
\begin{align}
\frac{\Gamma(n)}{(k^2+\mu ^2)^n} = \int_0^{\infty} d\tau \tau^{n-1} e^{- \tau (k^2+\mu ^2)} \rightarrow \nonumber\\
\rightarrow \int_{1/\Lambda^2}^{\infty}  d\tau \tau^{n-1} e^{- \tau (k^2+\mu ^2)}.
\end{align}
Thus it is trivial to obtain within the proper-time method that $A \neq B$ since
\be
A_{\Lambda} = \frac{-2 i}{(4 \pi)^2} (\Lambda^2 - \mu ^2 \ln \Lambda^2/\mu ^2),
\ee
whereas
\be
B_{\Lambda} = \frac{- i}{(4 \pi)^2} (\Lambda^2 - 2 \mu ^2 \ln \Lambda^2/\mu ^2).
\ee

%If we introduce $\Lambda$ as the maximum radius of a spherical integration in the Euclidean space to solve the integrals (\ref {3}) and (\ref{4}), 
%this already determines the constants as being $c_{1}= -1$ and $c_{2}= -1$.
%In the context 
%of the Higgs decay in two photons, ref. \cite{Piccinini} solves an UV logarithmic divergent integral using two different contours (spherical and 
%elliptical) and this leads to different values for the constant $c_{1}$. Then, they specify the contour of the integration on physical considerations, 
%i. e. choosing a contour which does not break gauge invariance. In our case, these physical considerations are Naturalness and consistency of scale symmetry breaking which are used to fix $c_1$ and $c_2$.

As we can see above, an inadequate regularization scheme could fix the constants $c_{1}$ and $c_{2}$ in a way that would lead us to inappropriate conclusions. In the context of this work, the value $c_{2}=-1$, obtained with a sharp cutoff regularization, would lead to the fine-tuning problem. In the context of scalar QED coupled to gravitation, a non-vanishing value of $c_{2}$ would make this theory asymptotically free \cite{Jean}.

\section{1-loop corrections to the Higgs propagator and theoretical bounds on the Higgs mass}\label {s3}

After spontaneous symmetry breaking, the contributions to the Higgs boson propagator include massive and Goldstone fields in the Landau gauge ($\xi =0$).

\begin{figure}[!htb]
 \centering
 \includegraphics[scale=0.8]{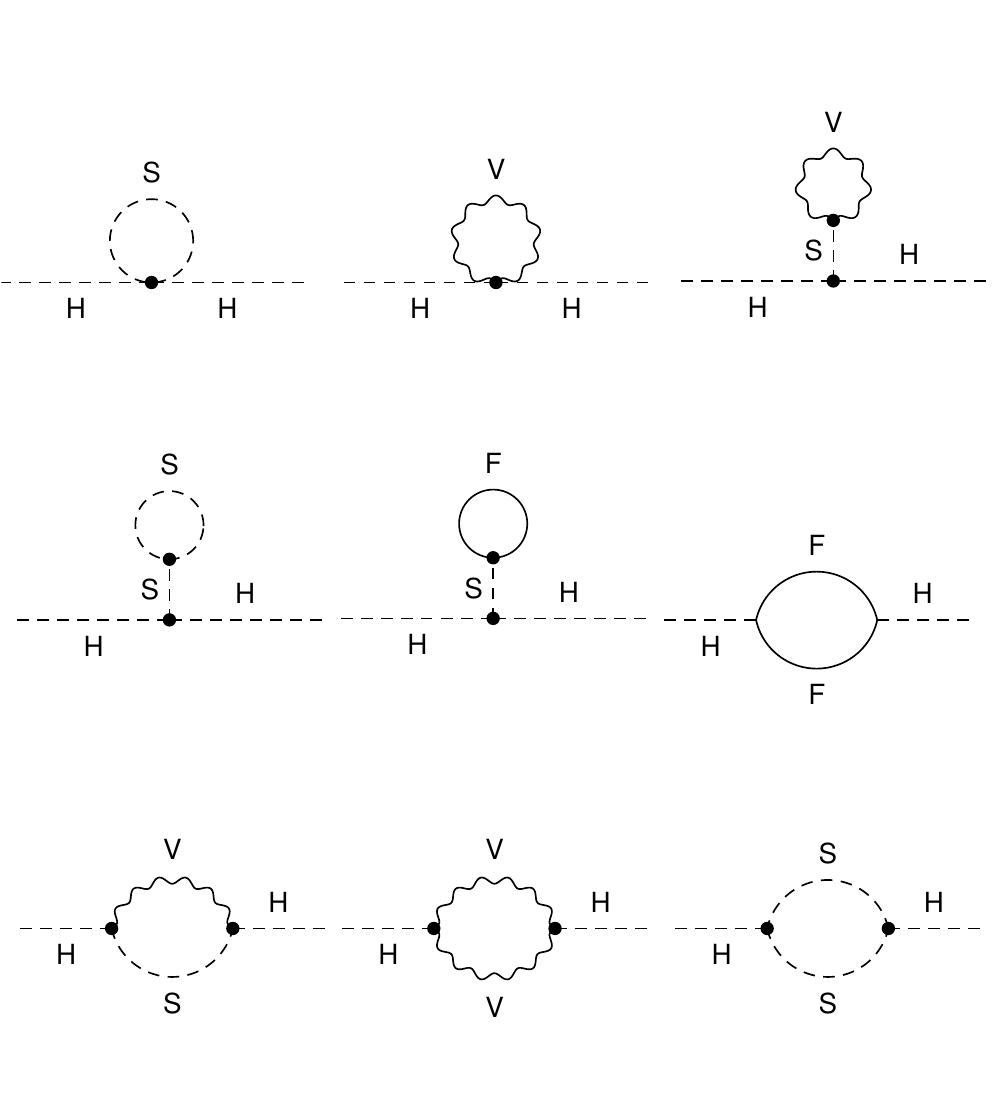}
 \caption{1-Loop correction to the Higgs propagator in the Landau gauge. S, V and F stands for scalar, vector boson and fermionic field, respectively.}
 \label{fig2}
\end{figure}

Figure \ref {fig2} shows all 1-loop contributions. We also consider only the heaviest fermion ( top quark) for the 
fermionic 1-loop diagrams. 

If one regularizes with a sharp covariant Euclidean cutoff 
and an on-mass-shell renormalization, for example, one has the Higgs mass correction given by equation (\ref {1}).  The quadratic 
cutoff of that equation means that we have an upper bound for this theory around 2 TeV \cite{{Chaichian},{Espinosa}}. This implies that the Standard Model, 
as an effective theory, should be valid up to energy scales that have already been reached by LHC \cite
{Espinosa}. Furthermore, one could expect the appearance of new physics around this energy, which so far did not emerge. So one can question whether the scale for the onset of new physics is really of that order. In addition, as the Higgs
mass is expected to be 125 GeV \cite{Higgs} we also would expect the cutoff to be much larger than that mass to preserve
the hierarchy of the theory.

In terms of the basic divergent integrals defined in eqs. \ref{8} and \ref{9}, the 1-loop contributions in figure \ref{fig2} can be written in an on-mass-shell renormalization, for example, as
%In contrast, if we use the implicit regularization scheme, where we do not assume a particular regularization scheme, for each 1-loop contribution of figure \ref {fig2} and the on-mass-shell
%regularization we have the Higgs mass correction given by
\begin{align}
&\delta m ^ 2 =\frac{6i}{\upsilon^2}[m_{Z}^2+2m_{W}^2+m^2-4m_{t}^2]I_{quad}(m ^2)+ \nonumber\\
&\frac{-3im^2}{\upsilon^2}[3m^2+6m^2_W+m^2_Z-6m^2_t]I_{log}(m ^2)
\label{10}
\end{align}
where we neglect finite terms in the UV limit because they are small compared to terms proportional to $\Lambda$ and
$\ln \Lambda$ when $\Lambda$ is large. For example, the finite integral in the UV limit of eq. (\ref {5}) is much smaller than one when $p^2= m^2_H$.

Now, considering eq. (\ref {9}), $\delta m^2$ correction reads
\begin{equation}
  \delta m ^ 2 = \frac{-3 c_{2}}{8 \pi^2 \upsilon^2}[m_{Z}^2+2m_{W}^2+m^2-4m_{t}^2]\Lambda ^2+... 
  \label{11}
\end{equation}
where the ellipses stand for contributions from $I_{log}(m ^2)$ and other terms from eq. (\ref {9}).

Because $c_{2}$ is arbitrary as discussed in section \ref{s2} it is consistent with a vanishing quadratic divergence. This assumption is justified by compatibility with scale
symmetry breaking which at one loop can be broken only by terms proportional to the beta function and terms which vanish in the limit $m\rightarrow0$ as argued in 
section \ref{s1}. Thus it consists of a theoretical symmetry argument that fix an arbitrariness in this case.

The constant $c_1$ from eqs. (\ref {8}) and (\ref{9}) is also chosen using this argument, i. e. the finite part in the UV 
limit of the diagrams can not contain terms which are not proportional to $m^2$ because they break scale invariance in the 
limit $m\rightarrow0$. These considerations complies with the definition of Naturalness imposed by 't Hooft \cite{Hooft}.

A second argument from the phenomenological standpoint in favor of this choice is to assure Naturalness of the SM and the consequent absence of the fine-tuning problem. This 
choice appears to be more appealing than the one made in refs. \cite{Veltman} and \cite{Castro}, since now the quadratic energy scale vanishes without the 
need to enforce a constraint among masses of different particles which may vary differently with the energy scale. 

There remain only the terms proportional to $m^2\ln\frac{\Lambda}{m}$ which do not break scale symmetry in the limit $m\rightarrow0$. These terms come from 
the integrals $I_{quad}(m ^2)$ and $I_{log}(m ^2)$ rewritten using eqs. (\ref {8}) and (\ref {9}). So, the correction to the Higgs mass is given by  
\begin{equation}
  \delta m ^ 2 = \frac{3 m^ 2_H}{16 \pi^2 \upsilon^2}[2m_{t}^2+2m_{W}^2+m^2_H-m_{Z}^2] \ln \frac {\Lambda^ 2}{m^2_H} +... 
  \label{12}
\end{equation}
where we have used $m^2_H=m^2+O(\frac{3m^2}{16\pi^2\upsilon^2})$ to replace $m$ by $m_H$. The neglected terms are of order of two-loop corrections.

If we consider only contributions of the order $\ln\frac{\Lambda}{m_H}$, we can use experimental data for the masses ($m_{t}=173 GeV$, $m_{W}= 80.2 
GeV$ and $m_{Z}= 91.2 GeV$), the vev value ($\upsilon=246\ GeV$) and the currently accepted value for the Higgs mass ( $m_H=125\ GeV$), to 
estimate the limit of validity of perturbation theory. To do this, we ask where perturbation theory starts to fail, i. e. where $\delta m= O(m_H)$. We 
get $\Lambda{\approx} 10^{10} GeV$. 

We can obtain a more refined estimate if we consider running masses in equation (\ref {12}). 
To include this information in that equation, we use the usual Standard Model relations between masses and coupling constants. Then we solve the beta functions of ref. \cite{Chaichian} in order to find the running coupling constants. We have
\begin{equation}
\frac{\delta m^2}{m^2_H}=\frac{3}{8\pi^2}[\frac{m^2_H}{\upsilon^2}+\frac{1}{4}g^2(\Lambda)-\frac{1}{4}g'^2(\Lambda) +g^2_t(\Lambda)]\ln\frac{\Lambda}{m_H}
\label{13}
\end{equation}
where $g(\Lambda)$, $g'(\Lambda)$ and $g_t(\Lambda)$ are the solutions of one loop beta functions for the SU(2), U(1) and Yukawa 
couplings, respectively. We use as initial values $g^2(m_Z)= 0.42$, $g'^2(m_Z)= 0.13$ and $g_t(m_t)= 1.01$. 

\begin{figure}[!ht]
\centering
    \subfigure[]
    {   \includegraphics[scale=0.8]{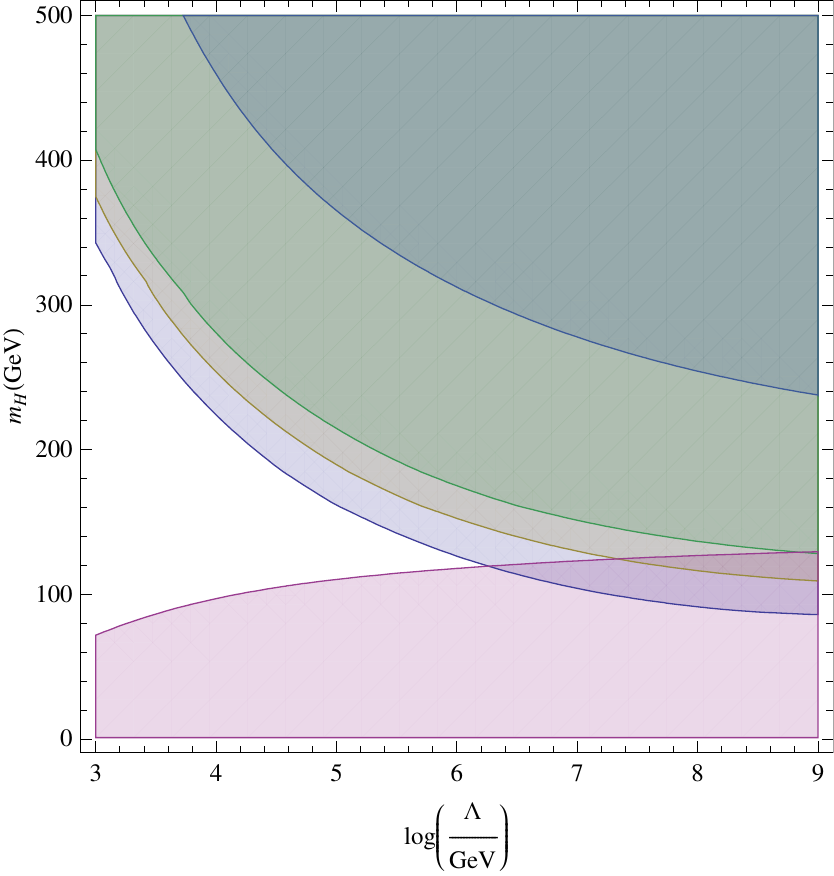}
        \label{fig3a} }
    \subfigure[]
{   \includegraphics[scale=0.8]{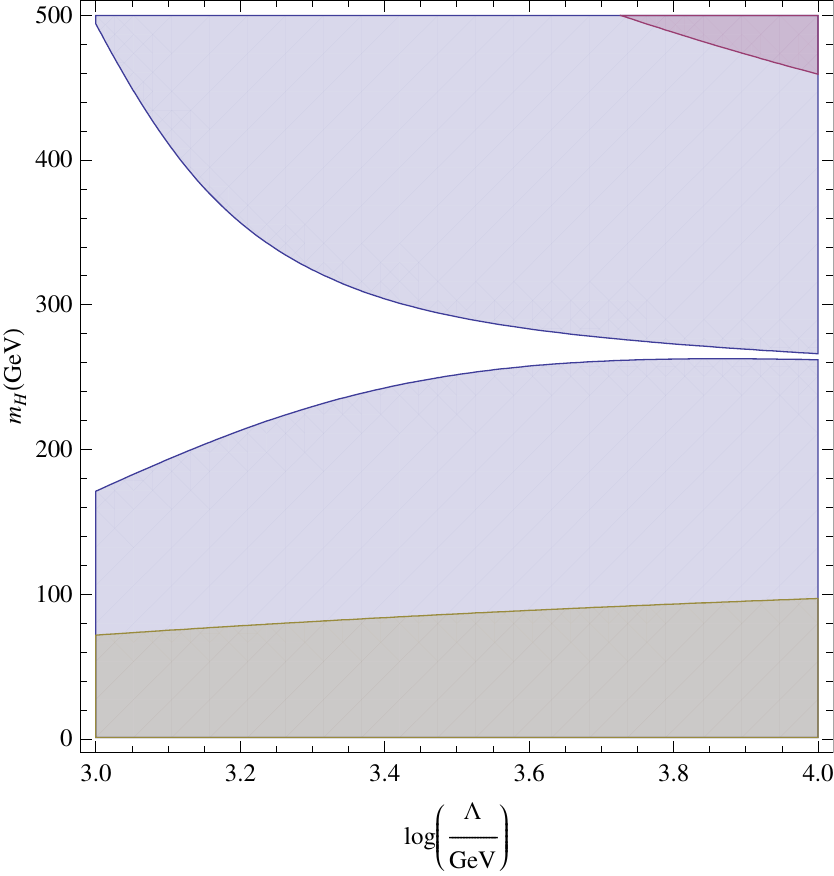}
        \label{fig3b}}
\caption{(a)Excluded values for the Higgs mass: the pink bound is the {\it vacuum stability bound} \cite{Regiao2} and the dark green one is the {\it triviality 
bound} \cite{Regiao1}. The blue, brown and light green bounds are the ones which exclude all values of $m_H$ and $\Lambda$ which lead to corrections $\delta m^2$ 
larger than or equal to $35\%$, $40\%$ and $45\%$ of $m^2_H$, respectively.(b)Excluded values for the Higgs boson mass considering $\Lambda^2$ dependence: the brown 
bound is the {\it vacuum stability bound} \cite{Regiao2} and the pink one is the {\it triviality bound} \cite{Regiao1}. The blue bound is the one which excludes all 
values of $m_H$ and $\Lambda$ which lead to corrections $\delta m^2$ larger than or equal to $m^2_H$.}
\label{fig3}
\centering
\end{figure}

The estimate now (for $\upsilon=246\ GeV$ and $m_H=125\ GeV$) is $\Lambda{\approx} 10^{16} GeV$. At this scale $\delta m$ starts to be equal or 
higher than $m_H$. That limit would be of order $1\ TeV$ should quadratic divergences be included \cite
{{Chaichian},{Espinosa}}.

Instead of assuming that perturbation theory remains valid up to the scale of $10^{16} GeV$ , one can alternatively present estimates for $\delta m^2/m^2_H$ in the range $0$ to $1$ where perturbation theory is still acceptable. A more complete and adequate analysis is to use eq. (\ref {13}) plus 
the finite part in the UV limit to exclude a set of values of $m_H$ and $\Lambda$ that yields values of corrections outside the $\delta m^2/m^2_H$ value we have chosen. The result is a region of excluded values of the Higgs mass in a certain energy scale.
Figure \ref {fig3a} shows a blue, a brown and a light green bound 
which excludes all values 
of $m_H$ and $\Lambda$ whose correction $\delta m^2$ is larger than or equal to $35\%$, $40\%$  and $45\%$ of the Higgs mass $m_H$, respectively. There 
is a pink bound which 
corresponds to the {\it vacuum stability bound} \cite{Regiao2} that excludes all values of $m_H$ and $\Lambda$ that lead to a negative Higgs 
self-coupling. Our bounds also coincide in part with the {\it triviality bound} \cite{{Regiao1}}(dark green bound) which excludes all values of $m_H$ 
and $\Lambda$ which trespass the Landau pole. For $\delta m^2/m^2_H\sim 0.1$ we obtain bounds which essentially overlap with the vacuum stability bound and
prevent the existence of allowed values for the Higgs mass. On the other hand, a bound $\delta m/m_H \rightarrow 1$ approaches the triviality bound and
large ranges for $m_H$ are obtained.

For a given value of the mass correction, we can determine a Higgs mass range or a Higgs mass value in a certain energy scale. The crossing of the {\it vacuum stability bound} with a $\delta m^2/m_H^2$ percentage curve yields the minimal mass correction that one can achieve at a certain scale. Its crossing with the blue mass region means that for a $35\%$ correction and in a energy around $10^3\ TeV$, the 
Higgs mass value is around $119\ GeV$. For the brown region, we have a Higgs mass 
value around $124\ GeV$ in an energy around $10^4\ TeV$, if the mass correction squared is around $40\%$. That means that if the cutoff of the SM
is around that order we still have a Higgs boson mass near the expected one \cite{Higgs}. The same interpretation can be given for the light green bound. In that 
case, we obtain a Higgs mass around $129 GeV$ for energies around $10^6\ TeV$. We see that the greater the correction value, the greater the allowed regions for the Higgs mass. For example, the crossing of the bound obtained for a $20\%$ correction with the
vacuum stability bound occurs at $10\ TeV$ and we have a smaller white region than the ones of figure \ref{fig3a}.

The use of a consistent scale symmetry breaking to fix an arbitrary parameter that controls the quadratic divergence excludes the fine-tuning problem and consequently, we have a reliable perturbation 
theory. The choices $\delta m^2<m^2_H$ are also in agreement with the Renormalization Group because those bounds do not trespass the {\it triviality bound}. 
In the scenario where fine-tuning is used, the Higgs mass should approach $160\ GeV$ at an energy of $100\ TeV$ \cite{Regiao3}. However,
the Higgs mass correction of ref. \cite{Regiao3} violates perturbation theory and the consistency of the scale symmetry breaking.

Furthermore, these bounds give another argument against the $\Lambda^2$ dependence. If that dependence exists, the measured value of $125 GeV$ is excluded for all energy scales and for all $\frac{\delta m^2}{m^2_H}<1$. We draw the bound corresponding to $\frac{\delta m^2}{m^2_H}=1$ (the maximum perturbation theory allows) considering the $\Lambda^2$ dependence in figure \ref{fig3b}. That means if we want the Higgs mass around the expected value we must allow some fine-tuning, i.e. $\delta m^2/m^2_H>1$ , as we can see in ref. \cite{Regiao3}. On the other hand, the result of figure \ref {fig3a} allows Higgs masses around the expected value for $\delta m^2/m^2_H<1$, where perturbation theory is acceptable.

\section {Concluding remarks}

We showed that regularization dependent arbitrariness which appear in quadratic divergences can be parametrized as shown in section \ref{s2} and fixed on symmetry grounds to restore Naturalness and perturbation theory for the Higgs boson mass in the SM. 
We have analyzed the quantum corrections to the Higgs boson mass together with the vacuum stability bound.
Although the correction values are not known, it is possible to establish bounds if we choose certain values for  $\delta m^2/m^2_H$. These bounds provide phenomenological arguments against the existence of $\Lambda^2$ dependence in the Higgs boson mass correction.

\vspace{0.5cm}

{\bf Acknowledgments}

Research supported by Funda\ca o para a Ci\^ecia e Tecnologia, 
project CERN/FP/116334/2010, developed under the iniciative QREN, financed by
UE/FEDER through COMPETE - Programa Operacional Factores de Competitividade. 
This research is part of the EU Research Infrastructure Integrating Activity 
Study of Strongly Interacting Matter (HadronPhysics3) under the 7th Framework 
Programme of EU: Grant Agreement No. 283286. 

The authors thank CNPq and FAPEMIG for financial support and Dr. Luis Cabral
for fruitful discussions.

\end{document}